\def\year{2016}
\documentclass[letterpaper]{article}
\usepackage{aaai16}
\usepackage{times}
\usepackage{helvet}
\usepackage{courier}
\usepackage{graphicx}
\usepackage{float}
\usepackage{url}
\usepackage{color}
\usepackage{multirow}
\usepackage{tabularx}
\usepackage[font=small,labelfont=bf,tableposition=top]{caption}

\begin{document}


\title{Toward Crowdsourced User Studies for Software Evaluation}

\author{Florian Daniel$^1$, Pavel Kucherbaev$^2$\\
$^1$ Politecnico di Milano, DEIB, Via Ponzio 34/5, Milano, Italy, 20133\\
florian.daniel@polimi.it\\
$^2$ University of Trento, DISI, Via Sommarive 9, Povo, Italy, 38123\\
pavel.kucherbaev@unitn.it\\
}

\makeatletter
\let\@oldmaketitle\@maketitle
\makeatother
\maketitle

\section{Goal}
This work-in-progress paper describes a \emph{vision}, i.e., that of \emph{fast 
and reliable software user experience studies conducted with the help from 
the crowd}. Commonly,
user studies are controlled in-lab activities that require the instruction, 
monitoring, interviewing and compensation of a number of participants that 
are typically hard to recruit. The goal of this work is to study which user study 
methods can instead be crowdsourced to generic audiences to enable the 
conduct of user studies without the need for expensive lab experiments. 
The challenge is understanding how to conduct crowdsourced studies
without giving up too many of the guarantees in-lab settings are able to
provide.

User studies are experimental and observational research methods for 
the measurement of an artifact's properties as perceived by its users 
(we specifically focus on software artifacts, such as Web applications). 
They are, for instance, used to evaluate the strengths and weaknesses 
of different visualization techniques, to understand if theoretical principles 
hold in practical settings, to measure if requirements are met by a given 
software design, or to validate and test usability. Over the last decades, 
user studies have increasingly found their way into software engineering 
practice, and today it is almost impossible to find successful applications 
that do not consider the perception of their users. Facebook and Google, 
for example, can rely on an unprecedented user base to test new features
on the fly and to adjust them according to observed performance or
preferences. The problem is that not everybody has access to such a
user base, e.g., because the own application has only a small target user
group or because the application is still under development.

Streamlining the necessary methods, involving the crowd, and providing 
user study support as a service while 
keeping study outputs reliable can thus make user studies significantly 
more accessible, to the benefit of everybody. The focus of this work is 
on how to crowdsource different user study methods conceptually and 
technically, i.e., on how to design effective tasks for user studies, gather 
and analyze data, guarantee quality and achieve representativeness. 
The question which method suits which research question is outside of 
its scope.

\section{A Success Story}
The idea for this proposal stems from a concrete experience with a 
crowdsourced user study \cite{RoyChowdhury2014} in which we used 
Amazon Mechanical Turk (MTurk, https://www.mturk.com) to validate 
the effectiveness of a research prototype. The prototype was a 
recommender system for a graphical mashup modeling environment 
(Yahoo! Pipes) that was able to provide modelers with on-the-fly 
recommendations of model patterns; if a recommendation was 
accepted, the respective pattern was automatically woven into the 
model under development. The study wanted to assess if recommending 
model patterns indeed provides benefits in terms of reduced modeling 
time, reduced number of user interactions, and reduced thinking time 
(time between two user interactions). For this purpose, it was necessary 
to enroll a number of participants, to split them into a control group 
(without recommendations) and a test group (with recommendations), 
to provide them with the recommender system (a plug-in of Yahoo! 
Pipes) and a modeling scenario to develop, to log 
their user interactions, to reward them for their work, to collect 
data, and to run a set of statistical hypothesis 
tests to verify the claims of the work. 

Despite the typical difficulties of crowdsourcing (e.g., worker selection, 
quality control, reward tuning, the implementation and provisioning of 
MTurk-external software for activity logging), we were 
able to enroll 30 participants, to validate the prototype in few days 
and to achieve statistical significance. The same prototype was also 
tested in two in-lab studies that confirmed the results obtained with the 
crowd \cite{RoyChowdhury2014}, incepting the idea of a more general 
theory of crowdsourced user studies.

\section{State of The Art}
A wealth of literature exists on user studies in the area of Human-Computer 
Interaction (HCI). For instance, Lazar et al. \shortcite{Lazar2010} provide an 
excellent introduction to HCI research methods, such as surveys, diaries, 
interviews, usability tests, etc. and the respective experiment designs and 
statistical data analysis requirements. Albert and Tullis \shortcite{Tullis2013} 
more specifically focus on measuring user experience and define a set of 
metrics, e.g., for performance measures, self-reporting, or comparative 
analyzes. In \cite{Albert2009}, the same authors elaborate on how to 
conduct remote, online user experience studies, while Brush et al. 
\shortcite{Brush2004} report that their participants (local and remote) even 
explicitly preferred remote follow-up studies over local ones. Bakshy et al. 
\shortcite{bakshy2014www} describe PlanOut, a domain-specific language
that separates experimental design and application logic in online experiments.

The idea of crowdsourcing user studies was discussed among the first 
by Kittur et al. \shortcite{Kittur2008} in 2008, soon after the emergence of the 
first crowdsourcing platforms. While the potential for crowdsourced user 
studies (in particular, surveys) was thus identified early, the authors 
however pointed out some crowd-specific limitations, such as the impossibility 
to make between-subject studies (it's typically not possible to control which
worker performs which task) and the difficulty to measure subjective or qualitative 
properties (there is no ground truth for opinions that can be used to 
assess correctness). Nevertheless, Difallah et al. \shortcite{Difallah2015}
report an increase of survey tasks published on MTurk in the last years, 
underlining the importance of this kind of method. 

As for the differences between in-lab and crowdsourced user studies, 
the success story described previously \cite{RoyChowdhury2014} 
(a complex study with different user measurements and the coordination 
of multiple study steps) produced similar development time, user 
interaction and thinking time measures across in-lab and crowdsourced 
experiments. Other researchers also compared these two types of experiment 
settings \cite{Heer2010,Liu2012,NebelingSN13a} concluding results in both 
settings with lower costs and more diversity in crowdsourced setting and 
some differences in demographics and user behavior between the two. 
Building on these experiences, understanding how to level out these differences 
is the object of this proposal. 


\section{The Research Questions}
The research questions that ask for answers to lay the foundation 
of principled, crowdsourced user studies are: 

\begin{itemize}

\item \emph{Which user study methods can be ported from the lab to the crowd?}
A wealth of user research methods exist, ranging from attitudinal,
qualitative methods like interviews and focus groups to behavioral, quantitative
methods like clickstream analysis and eye tracking. Behavioral, quantitative
methods can of course be automated more easily. The question is how much
crowdsourced user studied can be pushed into the area of attitudinal,
qualitative methods.

\item \emph{Which user experience questions can be answered reliably? }
A method may be used to answer one or more research questions. For
example, an interview can provide insight into preferences and values, while
A/B testing allows one to compare two features. If a method is crowdsourced, 
can it still answer the same questions? How do we guarantee
significance and representativeness?

\item \emph{Which assumptions, limitations and theories govern crowdsourced 
user studies?}
User studies assume, for instance, that participants can be 
recruited according to pre-defined profiles, an assumption that not necessarily
holds if the study is crowdsourced. If not, which implications does this
have on the expressiveness of the study?

\item \emph{Which software instruments can assist crowdsourced user studies?}
Crowdsourcing heavily leverages on micro-task platforms. Are these
suitable to support crowdsourced user studies? Or could dedicated extensions
or alternatives mitigate existing shortcomings (e.g., the impossibility to select
workers) or even open up new opportunities?

\item \emph{How can the data exposed by software artifacts be protected?}
While privacy is regulated by laws and technically easy today, the problem that
may prevent potential experimenters to crowdsource their study is the need to 
publish assets, e.g., data or graphical designs, that may still be under development
and/or require protection. How do we crowdsource a user study that guarantees
data protection?

\end{itemize}

We do not yet have answers to these questions -- in line with the vision style of
this submission. We have some anecdotal, positive experiences with own user 
studies of research prototypes and task designs and are sure the HCOMP audience,
too, has some to share. The intuition is that the crowd may 
indeed represent a viable alternative in some types of studies. Which ones, we
will try to answer systematically.

\small

\bibliographystyle{names}
\bibliography{references}

\bigskip

\end{document}